\documentclass[12pt]{article}
\usepackage{amsfonts}
\usepackage[dvips]{graphicx}
\usepackage{latexsym,amsmath,amssymb,graphics,stmaryrd}
\usepackage{cite}
\usepackage[bulletsep]{collref}
\usepackage{array}
\usepackage{subfigure}
\usepackage{multirow}
\usepackage{epsfig}
\usepackage[dvips]{graphicx}
\usepackage[dvips]{color}
\usepackage{pstricks}
\usepackage{pst-node}
\usepackage{pst-plot}
\usepackage{dsfont}
\usepackage{amsthm}
\usepackage{graphics}
\usepackage{subfigure}
\usepackage{fancyhdr}
\usepackage[dvips]{color}

\fancypagestyle{plain}{
\fancyhf{}
\newcommand{\fpage}{\iffloatpage{}{\thepage}}
\fancyfoot[C]{\fpage}

}

\newcommand{\be}{\begin{eqnarray}}
\newcommand{\ee}{\end{eqnarray}}

\newcommand{\la}{\lambda}

\newcommand{\NN}{\mathcal{N}}
\newcommand{\DD}{\mathcal{D}}
\newcommand{\ZZ}{\mathcal{Z}}
\newcommand{\CC}{\mathcal{C}}
\newcommand{\GG}{\mathcal{G}}
\newcommand{\OO}{\mathcal{O}}
\newcommand{\AAA}{\mathcal{A}}
\newcommand{\s}{\sigma}
\newcommand{\half}{\sfrac{1}{2}}
\newcommand{\al}{\alpha}
\newcommand{\p}{\partial}
\newcommand{\lb}{\langle}
\newcommand{\rb}{\rangle}
\newcommand{\tQ}{\widetilde Q}
\newcommand{\tS}{\widetilde S}
\newcommand{\dda}{{\dot{\alpha}}}
\newcommand{\ddb}{{\dot \beta}}

\newcommand{\de}{\delta}
\newcommand{\ve}{\varepsilon}

\newcommand{\tM}{{\tilde M}}
\newcommand{\tN}{{\tilde N}}
\newcommand{\tL}{{\tilde L}}
\newcommand{\slk}{\slash\!\!\!k}

\newcommand{\e}{\operatorname{e}}
\newcommand{\eps}{\varepsilon}
\newcommand{\Scl}{S_{\mbox{cl}}}

\newlength{\neglength}
\newlength{\diameter}
    
  \newcommand{\nn}{\nonumber}
  \newcommand{\sech}{{\,\rm sech}}
 \newcommand{\sfrac}[2]{\mbox{$\frac{#1}{#2}$}}

\numberwithin{equation}{section}
\newlength{\unit}
\psset{xunit=\unit,yunit=\unit,runit=\unit}
\newlength{\linew}
\psset{linewidth=\linew}
\begin{document}
%\immediate\write18{open -g -a spires \jobname.tex}

%\thispagestyle{empty}
\begin{flushright}\footnotesize
\texttt{UUITP-02/12} %\\
%\texttt{NBI-??}
\vspace{0.8cm}\end{flushright}

\begin{center}
{\Large\textbf{\mathversion{bold} Holographic three-point functions for short operators}}

\vspace{1.5cm}

\textrm{Joseph~A.~Minahan}
\vspace{8mm}

\textit{ Department of Physics and Astronomy\\ Uppsala University\\ Box 520\\
SE-751 20 Uppsala, Sweden}\\
\texttt{joseph.minahan@physics.uu.se} \vspace{3mm}

\vspace{3mm}

%%%%%%%%

\par\vspace{1cm}

\textbf{Abstract} \vspace{5mm}

\begin{minipage}{14cm}

We consider  holographic three-point functions for operators dual to short string states at strong coupling in $\NN=4$ super Yang-Mills.  We treat the states as point-like as they come in from the boundary but as strings in the interaction region in the bulk.  The interaction position is determined by saddle point,  which is equivalent to conservation of the canonical momentum for the interacting particles, and leads to conservation of their conformal charges.  We further show that for large dimensions the {\it rms} size of the interaction region  is small compared to the radius of curvature of the $AdS$ space, but still large compared to the string Compton wave-length.  Hence, one can approximate the string vertex operators as flat-space vertex operators with a definite momentum, which depends on the conformal and $R$-charges of the operator.  We then argue that the  string  vertex operator dual to  a primary operator is chosen by satisfying a twisted version of $Q^L=Q^R$, up to spurious terms.  This leads to a unique choice for a scalar vertex operator with the appropriate charges at the first massive level.  We then comment on some features of the corresponding three-point functions, including the application of  these results to  Konishi operators.

\end{minipage}

\end{center}

\vspace{0.5cm}

%%%%%%%%%%%%%%%%%%%%%%%%%%%%%%%%%%%%%%%%%%%%%%%%%%%%%%%%%%%%%%%%%%%%%%%%%%%%%%%

\newpage
\setcounter{page}{2}
\renewcommand{\thefootnote}{\arabic{footnote}}
\setcounter{footnote}{0}
\setcounter{equation}{0}

%%%%%%%%%%%%%%%%%%%%%%%%%%%%%%%%%%%%%%%%%%%%%%%%%%%%%%%%%%%%%%%%%%%%%%%%%%%%%%

\section{Introduction}

In this paper we will consider three-point functions for  ``short" operators at level one.
According to the AdS/CFT correspondence \cite{Maldacena:1997re,Gubser:1998bc,Witten:1998qj}, type IIB string theory on $AdS_5\times S^5$ is equivalent  to $\NN=4$ super Yang-Mills.  At large 't Hooft coupling $\la$, the radius of curvature of $AdS_5$ and  $S^5$ becomes large and the dimensions of  short operators can be accurately computed by assuming that the space is flat.  For example, there are expected to be operators with dimensions $\Delta\approx2\sqrt{n}\,\la^{1/4}$ corresponding to the energies of stringy modes when $\alpha'=1/\sqrt{\la}$ \cite{Gubser:1998bc}.  Taking advantage of integrability, it now is clear that at strong coupling the Konishi operator approaches one of these stringy modes with $n=1$ \cite{Bianchi:2003wx,Beisert:2003te,Gromov:2009zb}\nocollect{Gromov:2009zb}.

An interesting question is whether one can use the flat space limit to consider certain  higher point correlators.  Here, we limit ourselves to three-point correlators.    If we are in the flat space limit then the correlators should be related to stringy three-point functions using flat-space vertex operators.  At first glance this seems perplexing.  In flat space  there is energy conservation between the three string states and if we naively translate this over via AdS/CFT it would appear that we could only consider  correlators where the dimension of one operator is the sum of the dimensions of the other two.  Obviously there is no such requirement for three-point  correlation functions.  It must then be the case that the AdS curvature  plays a vital role.

In considering this problem we will make use of the results in  \cite{Janik:2010gc}.  Here the authors considered the path integral for three particles propagating in from the boundary of $AdS_d$ and intersecting at an intersection point in the bulk.  The intersection point is determined by minimizing the action for the three particles.  Here we will explicitly determine its location.  We will show that the solution can be thought of as a momentum conservation equation and these momenta are precisely what should go into the vertex operators.  We will further show that the fluctuations of the position of the vertex are over a sufficiently small volume of the bulk that one can treat the interaction region as being approximately a flat-space region.

Once in the flat-space region, we need to determine the appropriate vertex operators.  The operators under consideration are primary operators.  These are highest weight states in representations of the superconformal algebra, and as such are annihilated by all superconformal generators.  Taking the flat-space limit we will show that the primary operator condition is equivalent to setting the action of $Q^L$ acting on a  flat-space vertex operator equal to  $\pm\, Q^R$ acting on the same operator,   up to spurious terms, where the sign depends on the spin component.  This allows us to uniquely determine a scalar vertex operator at level one satisfying this criterion.    The $Q^L=\pm\, Q^R$ condition inevitably leads to a mixing between NS-NS and R-R modes for the vertex operators.

Using these vertex operators, one can then find  the three-point coupling to use in the correlators.  At level one, the combinatorics is messy and the final result will be presented in a separate publication \cite{To_appear}.  However, there is still much we can say about the behavior  of the the 3-point functions.   For example, one can  show that in the limit where the dimension of the operator $\Delta$ is much greater than its $R$-charge $J$,  the only dependence on $J$ is through overlap integrals and not in the vertex functions.  This suggests that one can take the limit of $J=0$, which is the case for the Konishi operator and still use the flat-space vertex operators, but using the $S^5$ overlap integrals.  We can also show that the correlator is exponentially suppressed in the limit of strong coupling for three primaries when  all satisfy $\Delta\gg J$. 

The results presented here are intended to complement many recent results in the subject of three-point functions.  These include the semi-classical results   for  two ``heavy" and one ``light" operator \cite{Zarembo:2010rr,Costa:2010rz,Roiban:2010fe,Hernandez:2010tg,Arnaudov:2010kk,Georgiou:2010an,Russo:2010bt,Park:2010vs,Buchbinder:2010ek,Arnaudov:2011wq,Bai:2011su,Alizadeh:2011yt,Michalcik:2011hh,Arnaudov:2012xx}\nocollect{Roiban:2010fe},  three heavy operators \cite{Ryang:2011tk,Janik:2011bd,Buchbinder:2011jr,Kazama:2011cp,Ryang:2012pm,Kazama:2012is},  BMN operators with large charges \cite{Klose:2011rm}, Wilson loops \cite{Alday:2011pf,Alday:2011ga,Hernandez:2012zj} and  various other configurations \cite{Bak:2011yy,Bissi:2011dc,Hernandez:2011up,Ahn:2011dq}.  
At weak coupling there has also been some significant results using the Bethe ansatz \cite{Escobedo:2010xs,Escobedo:2011xw,Gromov:2011jh,Foda:2011rr,Gromov:2012uv,Foda:2012wf,Bissi:2011ha,Georgiou:2012zj,Gromov:2012vu,Kostov:2012jr,Serban:2012dr,Kostov:2012yq}\nocollect{Bissi:2011ha}.  There are also results showing matching between the weak and  strong coupling limits for certain operators \cite{Georgiou:2011qk,Caetano:2011eb,Grignani:2012yu,Grignani:2012ur}.

In section 2 we review the construction of \cite{Janik:2010gc} for finding two and three-point functions for operators using the path integral for particles in an $AdS_{d+1}$ background.    For the three-point functions we find  the exact configuration that extremizes the action, showing that it is equivalent to a conservation of  the canonical momentum in $AdS_{d+1}$ space.  We  find the three-point correlator for three scalar operators, including   Gaussian fluctuation pre-factors, up to an overall coupling term.  We show that the size of the fluctuations are large compared to $\Delta^{-1}$, but small compared to the radius of $AdS_{d+1}$.

In section 3 we discuss how conserved charges of the conformal symmetry translate into conserved quantities for the path integral.  In section 4 we include the $S^5$ background in the path integral.  In section 5 we discuss the flattening of the conformal and superconformal algebras into Poincar\'e and superPoincar\'e algebras.  In particular, we show that setting the superconformal generators to zero is equivalent to setting $Q^L=\pm \,Q^R$ for the flattened superPoincar\'e algebra.   We then find the scalar vertex operators that satisfy this condition at level 0 and level 1.  We do this by first finding  vertex operators that satisfy $Q^L=Q^R$ and then twisting the result.  In section 6 we discuss some properties of the three-point coupling that can be used to infer behavior for the correlators.  In section 7 we present our conclusions.

\medskip
\noindent{\it Note:}  The results of section 2 were derived prior to the appearance of \cite{Klose:2011rm}\footnote{I thank T. Klose and T. McLoughlin for kindly acknowledging this in their own publication.}.

\medskip
\noindent{\it Note added:}  During the preparation of this manuscript \cite{Buchbinder:2011jr} appeared which has some overlap with the results of section 4.

\section{Review and extension of previous results}

In this section we review the analysis in \cite{Janik:2010gc} to compute 3-point correlators from a path integral.  We will also extend these results to show how to derive the semiclassical correlator for three arbitrary chiral primaries, each with dimension $\Delta_i\gg 1$.

We need to first consider the two-point correlator.  Suppose we consider the Euclidean path integral for a particle in $AdS_{d+1}$, dual to a CFT operator with dimension $\Delta$.  The action is given by \cite{Janik:2010gc}
\be\label{action}
S=\frac{1}{2}\int_{-1}^{+1} ds\left[\frac{\dot{  x}^\mu \dot{ x}_\mu+\dot z^2}{z^2} e^{-1}+\Delta^2e\right]
\ee
where $ x^\mu$ are the flat coordinates in $R^d$ and $e(s)$ is an {\it einbein}. Notice that the second term comes with a factor of $\Delta^2$ and not $m^2=\Delta(\Delta-d)$.  This is because $m^2$ is the square Casimir of the representation of the conformal group and hence we would not expect it to appear directly in the action\footnote{Even though we are considering large $\Delta$, we are also interested in the prefactors where the form of the second term will make a difference.}. Without any loss of generality we can assume that the endpoints at $s=\pm 1$ are along the $x^1\equiv x$ direction, with $x(\pm1)=\pm x_0,$, $z(\pm1)=\eps$.
The action is invariant under reparameterizations $s\to s'$, with $e'(s') =\left(\frac{\partial s'}{\partial s}\right)^{-1} e(s)$.  If we assume that the endpoints after reparameterization are still at $\pm1$, then the {\it einbein} can always be set to a constant $e'(s')=E$, where the new coordinate is given by
\be
s'=\frac{1}{E}\int_{-1}^s e(s)ds-1\,,
\ee
Thus $E$ is determined from $e(s)$,
\be
E=\frac{1}{2}\int_{-1}^{+1} e(s) ds\,.
\ee
and one can see that different values of $E$ correspond to inequivalent gauge classes.  Thus the path integral itself has the form
\be
\ZZ=\int \prod_{s=-1}^{s=1}\frac{\mu\, \DD  x^\mu(s)\DD z(s)\DD\, e(s)}{V_{diff}}\, e^{-S}=\int dE{\prod_{s=-1}^{s=1}\mu\, \DD  x^\mu(s)\DD z(s)}(Jac)\,e^{-S}
\ee
where $\mu$ is a measure factor, $V_{diff}$ is the gauge volume of the one-dimensional diffeomorphisms,  and $Jac$ is the Jacobian (Faddeev-Popov factor) that appears after gauge fixing, and for the most part will drop out, with one important exception.

The action is minimized by the solution \cite{Janik:2010gc}
\be\label{xzsol}
x(s)&=&R\tanh(\kappa s)\nn\\
z(s)&=&R\sech(\kappa s)\nn\\
 x^\mu_\perp&=&0
\ee
where $R\sech(\kappa)=\eps$ and $ x^\mu_\perp$ are the transverse components to $x^1$.  If $x_0\gg\eps$ then $\kappa\gg1$, $R\approx x_0$ and $\kappa\approx \log\left(\frac{2x_0}{\eps}\right)$\,.  Plugging (\ref{xzsol}) back into the action one finds
\be
S_{\mbox{cl}}=\kappa^2 E^{-1}+\Delta^2 E\,,
\ee
and so minimizing with respect to $E$ gives  $E=\kappa/\Delta$ and $S=2\kappa\Delta$.  Hence, the leading behavior of the path integral is \cite{Janik:2010gc}
\be
\ZZ\sim \Big|\frac{2x_0}{\eps}\Big|^{-2\Delta}\,,
\ee
which is the expected behavior for a two point correlator of a dimension $\Delta$  operator and its adjoint separated by $2x_0$.

In anticipation of the three-point correlators, we will redo the two-point amplitude slightly differently.  Let us break up the path integral into two parts where the particle path passes through the ``joining" point $( x^\mu,z)$ at $s=s_0$.  The path integral now has the form
\be\label{2point}
&&\int \left(dE_-{\prod_{s<s_0}\mu\, \DD  x^\mu(s)\DD z(s)}(Jac)\right)\frac{z^{-d-1}d^dx\, dz}{V_{Gauge}} \left(dE_+{\prod_{s>s_0}\mu\, \DD  x^\mu(s)\DD z(s)}(Jac)\right)\,e^{-S}\nn\\
&&=\int\frac{z^{-d-1}d^dx\, dz}{V_{Gauge}}K(\eps,+x_0^\mu;z,x^\mu)K(\eps,-x_0^\mu;z,x^\mu)\,,
\ee
where we  divided out by a residual gauge degree of freedom that enters because of the extra integral over an {\it einbein} component. We have also included the specific measure factor in the integral over $ x^\mu$ and $z$.  $K_\Delta(\eps,\pm x_0^\mu;z,x^\mu)$ are  boundary to bulk propagators \cite{Gubser:1998bc,Witten:1998qj} for a scalar field dual to an operator with dimension $\Delta$.  Minimizing with respect to $x(s)$, $z(s)$, $E_-$ and $E_+$ but keeping  $x(s_0)=x$ and $z(s_0)=z$, the action becomes
\be\label{S2pt}
\Scl=-\Delta\log\left(\frac{z\,\eps}{z^2+(x-x_0)^2+x_\perp^2}\right)-\Delta\log\left(\frac{z\,\eps}{z^2+(x+x_0)^2+x_\perp^2}\right)\,.
\ee
 Minimizing with respect to $ x^\mu$ and $z$ gives one of the points on the trajectory in (\ref{xzsol}).

Let us now consider the fluctuations in $ x^\mu$ and $z$.  For fluctuations in the transverse directions, we note that
\be
\frac{\partial^2 \Scl}{\partial x_\perp^\mu\partial x_\perp^\nu}=\left(\frac{2\Delta}{z^2+(x-x_0)^2}+\frac{2\Delta}{z^2+(x+x_0)^2}\right)\delta_{\mu\nu}=\frac{2\Delta}{z^2}\delta_{\mu\nu}\,.
\ee
Hence the integrals over the $d-1$ transverse components are approximately
\be
\int \frac{d^{d-1}x_\perp}{z^{d-1}}e^{-\Delta x_\perp^2/z^2}=\left(\frac{\pi}{\Delta}\right)^{\frac{d-1}{2}}\,.
\ee
For the longitudinal components $x$ and $z$, it is convenient to consider two new variables $\eta$ and $\xi$, where
\be
d\eta=\frac{zdz+xdx}{z\sqrt{z^2+x^2}},\qquad d\xi=\frac{zdx-xdz}{z\sqrt{z^2+x^2}}\,.
\ee
It is straightforward to check that $z^{-2}dx\,dz=d\eta\, d\xi$ and 
\be
\frac{\partial }{\partial\eta}=\frac{z}{\sqrt{z^2+x^2}}\left(z\frac{\partial}{\partial z}+x\frac{\partial}{\partial x}\right)\qquad\frac{\partial }{\partial\xi}=\frac{z}{\sqrt{z^2+x^2}}\left(z\frac{\partial}{\partial x}-x\frac{\partial}{\partial z}\right)\,.
\ee
One can then easily show that
\be
\frac{\partial^2}{\partial \eta^2}\Scl=2\Delta\,,\qquad \frac{\partial^2}{\partial \xi^2}\Scl=\frac{\partial^2}{\partial \xi\partial\eta}\Scl=0\,.
\ee
Hence, the integral over $\eta$ gives another factor of $(\pi/\Delta)^{1/2}$.  

The $\xi$ component is a zero mode.  Of course, the existence of the zero-mode is not surprising; it is  a consequence of the residual gauge invariance.  Let us now gauge fix such that $s_0$ is the corresponding value in (\ref{xzsol}) that gives $x$ and $z$.  This then fixes $\xi$.   One can easily show that $d\xi=\kappa ds$ and since the measure factor for the gauge volume is $E\,ds$, the ratio of the measures is 
\be
\frac{d\xi}{E\,ds}=\frac{\kappa}{E}=\Delta\,.
\ee
 Hence, the complete factor $B$ coming from the integral of the components at $s_0$ is
\be\label{Bval}
B=\frac{\pi^{d/2}}{\Delta^{d/2-1}}\,.
\ee
The fluctuation modes for $s<s_0$ and $s>s_0$ lead to overall factors $\CC_-$ and $\CC_+$ respectively.  The  path integral is then
\be
\ZZ\approx \CC_-\,\frac{\pi^{d/2}}{\Delta^{d/2-1}}\, \CC_+ \Big|\frac{2x_0}{\eps}\Big|^{-2\Delta}\,.
\ee

It is instructive to compare this formulation of the two-point function with the known results, which will lead to a correction factor.  If we take the joining point to the boundary, then  (\ref{2point}) becomes
\be
&&\frac{1}{\eps^{d-1}}\int{d^dx}{}\,K(\eps,+x_0^\mu;\eps,+x_0^\mu)\,\frac{\Delta}{z}\,K(\eps,-x_0^\mu;\eps,x_0^\mu)\nn\\
&&\quad=
\frac{1}{\eps^{d-1}}\int{d^dx}{}\,K(\eps,+x_0^\mu;\eps,+x_0^\mu)\,\partial_z K(\eps,-x_0^\mu;z,x_0^\mu)\Bigg|_{z=\eps}\,,
\ee
which is the expression found in \cite{Freedman:1998tz}.  However, it was further shown in  \cite{Freedman:1998tz} that this expression for the two-point function is not quite consistent with a Ward identity.  A different formulation was proposed that is consistent with the Ward identity \cite{Freedman:1998tz,Klebanov:1999tb} and leads to an overall correction factor of $\frac{2\Delta-d}{\Delta}$.  Since in general we assume that $\Delta\gg1$, we  will insert $2$ as the correction factor and use the corrected path integral
\be\label{Zcorr}
\ZZ_{corr}\approx \CC_-\,\frac{2\,\pi^{d/2}}{\Delta^{d/2-1}}\, \CC_+ \Big|\frac{2x_0}{\eps}\Big|^{-2\Delta}\,.
\ee

We now turn to the three-point correlator.  In this case we can treat this as a set of three half-propagators joining at an interaction point $( x^\mu,z)$\cite{Janik:2010gc}.  The interaction point is a preferred point on each of the propagators and does not change under a reparameterization.  Let us suppose that the three operators have dimension $\Delta_i$ and are located at the points $ x^\mu_i$.    From the discussion  of the two point correlators, it is clear that the minimized action has the form \cite{Janik:2010gc}
\be\label{Scl}
S_{\mbox{cl}}( x^\mu,z)=-\sum_{i=1}^3 \Delta_i\log\left(\frac{z\,\eps}{z^2+(x-x_i)^2}\right)\,.
\ee
If we minimize $S_{\mbox{cl}}( x^\mu,z)$ with respect to $ x^\mu$ and $z$ we end up with the set of equations
\be\label{3pteom}
\sum_i\frac{\Delta_i( x^\mu- x^\mu_i)}{z^2+(x_i-x)^2}&=&0\nn\\
\sum_i\frac{\Delta_i((x-x_i)^2/z-z)}{z^2+(x_i-x)^2}&=&0\,.
\ee

To get a better understanding of  the equations in (\ref{3pteom}), let us return to the equations of motion for the coordinates on the half-propagators.   Without any loss of generality we can write the solutions as
\be
 x^\mu(s)&=& R^\mu_i\tanh(\kappa_is+\beta_i)+ x^\mu_{0i} \nn\\
z(s)&=&R_i\sech(\kappa_is+\beta_i)\,,
\ee
where $s=0$ is the joining point, $s=-1$ is the end of the propagator at the boundary and $R_i=( R^\mu_i  R_{ \mu i})^{1/2}$.  From the positions of the endpoints we also have that
\be
 x^\mu_i&=& R^\mu_i\tanh(-\kappa_i+\beta_i)+ x^\mu_{0i}\approx  -R^\mu_i+ x^\mu_{0i}\qquad\qquad  x^\mu= R^\mu_i\tanh(\beta_i)+ x^\mu_{0i}\nn\\
\eps&=&R_i\sech(-\kappa_i+\beta_i)\approx 2R_i\,e^{-\kappa_i-\beta_i}\qquad\qquad\qquad z=R_i\sech(\beta_i)\,,
\ee
after which it follows that
\be\label{Riexp}
\frac{ x^\mu_i- x^\mu}{z^2+(x_i-x)^2}&\underset{\eps\to0}{=}&-\frac12\,\frac{ R^\mu_i}{R_i^2}\nn\\
\frac{(x_i-x)^2/z-z}{z^2+(x_i-x)^2}&\underset{\eps\to0}{=}&-\frac{\tanh(\beta_i)}{R_i\sech(\beta_i)}\,.
\ee
The $d+1$ dimensional canonical momentum along each trajectory is $\Pi_{m\,i}=( \Pi_{\mu i},\Pi_{z\,i})$, where
\be
\Pi_{\mu\,i}&=&-i\,E_i^{-1}\frac{\dot { x}_\mu}{z^2}=-i\,\Delta_i\frac{ R_{\mu i}}{R_i^2}\nn\\
\Pi_{z\,i}&=&-i\,E_i^{-1}\frac{\dot z}{z^2}=i\,\Delta_i\frac{\tanh(\kappa_i s+\beta_i)}{R_i\sech(\kappa_i s+\beta_i)}\,.
\ee
There is an overall factor of $i$ because the parameter $s$ is a Euclidean variable.
Hence, we see that the equations in (\ref{3pteom}) are nothing more than the conservation of all components of the canonical momentum at the joining point.

Furthermore, the  canonical momentum for each particle satisfies
\be\label{momsq}
\Pi_i\cdot\Pi_i\equiv G^{mn}\Pi_{m\,i}\Pi_{n\,i}=-z^2\Delta_i^2\left(\frac{1}{R_i^2}+\frac{\tanh^2(\kappa_is+\beta_i)}{R_i^2\sech^2(\kappa_is+\beta_i)}\right)=-\Delta_i^2\,.
\ee
Therefore, using the conservation laws at the joining point  we have that
\be
\Pi_1\cdot\Pi_2=\sfrac{1}{2}\left(\Delta_1^2+\Delta_2^2-\Delta_3^2\right)
\ee
with similar expressions for the other combinations.

It is tedious to derive but straightforward to confirm that the equations in (\ref{3pteom}) are satisfied by
\be\label{xzsols}
 x^\mu&=& x^\mu_1-\frac{\al_1}{F}\left(\al_2\,x_{12}^2\, x^\mu_{13}+\al_3\,x_{13}^2\, x^\mu_{12}\right)\nn\\
z^2&=&\frac{\al_1\al_2\al_3\Sigma}{F^2} x_{12}^2\,x_{13}^2\,x_{23}^2\,,
\ee
where
\be
\al_1=\half(\Delta_2+\Delta_3-\Delta_1)\,, &&\al_2=\half(\Delta_3+\Delta_1-\Delta_2)\,,\ \ \ \ \al_3=\half(\Delta_1+\Delta_2-\Delta_3)\,,\nn\\
\Sigma&=&\half(\Delta_1+\Delta_2+\Delta_3)=\al_1+\al_2+\al_3\,,\nn\\
F&=&\al_1\al_2\, x_{12}^2+\al_2\al_3\,x_{23}^2+\al_3\al_1\,x_{13}^2\,.
\ee
One can further show that the expression for $ x^\mu$ is consistent under interchange of the indices.  Alternatively, one can choose the more symmetric, although not necessarily  more convenient form
\be
 x^\mu=\frac{2}{3}( x^\mu_1+x^\mu_2+x^\mu_3)+\frac{1}{F}\left(\al_1\al_2x_{12}^2x^\mu_3+\al_2\al_3x_{23}^2x^\mu_1+\al_3\al_1x_{31}^2x^\mu_2\right)\,.
\ee
With the solutions in  (\ref{xzsols}), we have that
\be\label{x-xi}
\Delta_1\frac{x^\mu_1-x^\mu}{z^2+(x_1-x)^2}&=&\al_2\,\frac{x^\mu_{13}}{x_{13}^2}+\al_3\,\frac{x^\mu_{12}}{x_{12}^2}\nn\\
\Delta_2\frac{x^\mu_2-x^\mu}{z^2+(x_2-x)^2}&=&\al_3\,\frac{x^\mu_{21}}{x_{21}^2}+\al_1\,\frac{x^\mu_{23}}{x_{23}^2}\nn\\
\Delta_3\frac{x^\mu_3-x^\mu}{z^2+(x_3-x)^2}&=&\al_1\,\frac{x^\mu_{32}}{x_{32}^2}+\al_2\,\frac{x^\mu_{31}}{x_{31}^2}\,,
\ee
where the sum of the three terms is clearly zero, satisfying the first equation in (\ref{xzsols}).  Likewise, one can check that the solutions in (\ref{xzsols}) give
\be\label{useful}
\sum_i\frac{\Delta_i}{z^2+(x-x_i)^2}=\frac{\Sigma}{z^2}\,,
\ee
from which it follows that the second equation in (\ref{3pteom}) is satisfied.

Inserting the solutions in (\ref{xzsols}) back into $S_{\mbox{cl}}(x^\mu,z)$ in (\ref{Scl}) one finds
\be
S_{\mbox{cl}}&=&-\Delta_1\log\left(\frac{\al_2\al_3|x_{23}|}{\Delta_1|x_{12}||x_{13}|}\right)-\Delta_2\log\left(\frac{\al_3\al_1|x_{31}|}{\Delta_2|x_{23}||x_{21}|}\right)-\Delta_3\log\left(\frac{\al_1\al_2|x_{12}|}{\Delta_3|x_{31}||x_{32}|}\right)\nn\\
&=&\log\left(|x_{12}|^{2\al_3}|x_{23}|^{2\al_1}|x_{31}|^{2\al_2}\right)-\sum_i(\al_i\log\al_i-\Delta_i\log\Delta_i)-\Sigma\log\Sigma\,.
\ee
The $x_i$ dependence was previously derived in \cite{hep-th/9804058} and is the expected behavior for the three-point correlator of operators with dimensions $\Delta_i$.  

We are also interested in the contribution coming from the fluctuations over $x^\mu$ and $z$.  To this end, we note that 
\be
\frac{\partial^2}{\partial z^2}S_{\mbox{cl}}(x^\mu,z)&=&\sum_i\Delta_i\left(\frac{1}{z^2}-2\frac{z^2-(x-x_i)^2}{(z^2+(x-x_i)^2)^2}\right)\nn\\
&=&\frac{4\Sigma}{z^2}-4z^2\sum_i\frac{\Delta_i}{(z^2+(x-x_i)^2)^2}\nn
\ee
\be
\frac{\partial^2}{\partial x^\mu \partial x^\nu}S_{\mbox{cl}}(x^\mu,z)&=&\sum_i\Delta_i\left(2\frac{\delta_{\mu\nu}}{z^2+(x-x_i)^2}-4\frac{(x-x_{i})_\mu(x-x_{i})_\nu}{(z^2+(x-x_i)^2)^2}\right)\nn\\
&=&\frac{2\Sigma}{z^2}\,\delta_{\mu\nu}-4\sum_i\Delta_i\frac{(x-x_{i})_\mu(x-x_{i})_\nu}{(z^2+(x-x_i)^2)^2}\nn\\
\frac{\partial^2}{\partial x^\mu \partial z}S_{\mbox{cl}}(x^\mu,z)&=&-4\sum_i\Delta_i\frac{z\,(x-x_{i})_\mu}{(z^2+(x-x_i)^2)^2}
\ee
We further observe that all $x^\mu_i$ as well as $x^\mu$ lie in a plane.  Hence, without any loss of generality we may set $x_i^{m}=x^{m}=0$, $m=3\dots d$.  One can then show that
\be\label{det}
\det\left[\left(\begin{array}{ccc}\frac{\partial^2}{\partial z^2}&\frac{\partial^2}{ \partial z\partial x^1}&\frac{\partial^2}{ \partial z\partial x^2}\\
\frac{\partial^2}{\partial x^1 \partial z}&\frac{\partial^2}{\partial x^1 \partial x^1}& \frac{\partial^2}{\partial x^1 \partial x^2}\\
\frac{\partial^2}{\partial x^2 \partial z}& \frac{\partial^2}{\partial x^2 \partial x^1}& \frac{\partial^2}{\partial x^2 \partial x^2}
\end{array}\right)S_{\mbox{cl}}\right]=\frac{16}{z^6}\,\frac{\al_1\al_2\al_3}{\Delta_1\Delta_2\Delta_3}\,\Sigma^3\,.
\ee
Useful relations in deriving this result are (\ref{useful})
and 
\be
&&x_{12}^2x_{23}^2-\left(x_{12}\cdot x_{23}\right)^2=x_{23}^2x_{31}^2-\left(x_{23}\cdot x_{31}\right)^2=x_{31}^2x_{12}^2-\left(x_{31}\cdot x_{21}\right)^2\nn\\
&&\qquad=-\frac{1}{4}\left(x_{12}^4+x_{23}^4+x_{31}^4-2\,x_{12}^2x_{23}^2-2\,x_{23}^2x_{31}^2-2\,x_{31}^2x_{12}^2\right)\,.
\ee
Therefore, integrating over $x^\mu$ and $z$ with measure factor $z^{-d-1}d^dx\, dz$ gives the prefactor
\be
\frac{\pi^{\frac{d+1}{2}}}{\sqrt{2}}\left(\frac{\Delta_1\Delta_2\Delta_3}{\al_1\al_2\al_3\Sigma^{d+1}}\right)^{1/2}\,.
\ee

Assuming a coupling factor $\GG_{123}$, the complete path integral for the three-point joining is
\be
\ZZ_{123}\approx\CC_{1-}\CC_{2-}\CC_{3-}\,\frac{\pi^{\frac{d+1}{2}}}{\sqrt{2}}\,\left(\frac{\Delta_1\Delta_2\Delta_3}{\al_1\al_2\al_3\Sigma^{d+1}}\right)^{1/2}\,\frac{\al_1^{\al_1}\al_2^{\al_2}\al_3^{\al_3}\Sigma^\Sigma}{\Delta_1^{\Delta_1}\Delta_2^{\Delta_2}\Delta_3^{\Delta_3}}|x_{12}|^{-2\al_3}|x_{23}|^{-2\al_1}|x_{31}|^{-2\al_2}\GG_{123}\,.\nn\\
\ee
The factors $\CC_{i-}$ come from the fluctuations along the half-propagators.  Hence, using the corrected two-point function in (\ref{Zcorr}) the normalized correlator is
\be\label{3corr1}
\langle\OO_{\Delta_1}(x^\mu_1)\OO_{\Delta_2}(x^\mu_2)\OO_{\Delta_3}(x^\mu_3)\rangle&=&\frac{1}{\CC_{1-}\CC_{2-}\CC_{3-}}\frac{(\Delta_1\Delta_2\Delta_3)^{\frac{d-2}{4}}}{2^{3/2}\,\pi^{\frac{3d}{4}}}\,\ZZ_{123}\nn\\
&=&\frac{ C_{123} }{|x_{12}|^{\Delta_1+\Delta_2-\Delta_3}|x_{23}|^{\Delta_2+\Delta_3-\Delta_1}|x_{31}|^{\Delta_3+\Delta_1-\Delta_2}}\,,
\ee
where
\be\label{3corr2}
C_{123}\approx\frac{\pi^{\frac{2-d}{4}}}{4}\,\frac{(\Delta_1\Delta_2\Delta_3)^{d/4}}{\left(\al_1\al_2\al_3\Sigma^{d+1}\right)^{1/2}}\,\frac{\al_1^{\al_1}\al_2^{\al_2}\al_3^{\al_3}\Sigma^\Sigma}{\Delta_1^{\Delta_1}\Delta_2^{\Delta_2}\Delta_3^{\Delta_3}}\,\GG_{123}\,.
\ee

We can compare this to the supergravity calculation in \cite{Freedman:1998tz}.  For the two-point function for an operator dual to a bulk scalar field it was found that 
\be\label{sugra2pt}
\langle\OO_{\Delta}(x^\mu)\OO_\Delta(\vec y)\rangle=\frac{2\Delta-d}{\pi^{d/2}}\,\frac{\Gamma(\Delta)}{\Gamma(\Delta-\frac{d}{2})}\,\frac{1}{|x-y|^{2\Delta}}\approx\frac{2\,\Delta^{\frac{d+2}{2}}}{\pi^{\frac{d}{2}}}\,\frac{1}{|x-y|^{2\Delta}}\,,
\ee
We have absorbed all overall coefficients such that the scalar kinetic term is canonically normalized.  Likewise, the three-point function for three operators dual to bulk scalars was found to be
\be\label{sugra3pt}
 \langle\OO_{\Delta_1}(x^\mu_1)\OO_{\Delta_2}(x^\mu_2)\OO_{\Delta_3}(x^\mu_3)
=
\,\frac{\AAA_{123}}{|x_{12}|^{\Delta_1+\Delta_2-\Delta_3}|x_{23}|^{\Delta_2+\Delta_3-\Delta_1}|x_{31}|^{\Delta_3+\Delta_1-\Delta_2}}\,,
\ee
where
\be
\AAA_{123}&=&\frac{1}{2\,\pi^{d}}\,\frac{\Gamma(\al_1)\Gamma(\al_2)\Gamma(\al_3)\Gamma(\Sigma-\frac{d}{2})}{\Gamma(\Delta_1-\frac{d}{2})\Gamma(\Delta_2-\frac{d}{2})\Gamma(\Delta_3-\frac{d}{2})}\,\GG_{123}\nn\\
&\approx& 
\frac{\pi^{\frac{2-4d}{4}}}{\sqrt{2}}\,\frac{(\Delta_1\Delta_2\Delta_3)^{\frac{d+1}{2}}}{\left(\al_1\al_2\al_3\Sigma^{d+1}\right)^{1/2}}\,\frac{\al_1^{\al_1}\al_2^{\al_2}\al_3^{\al_3}\Sigma^\Sigma}{\Delta_1^{\Delta_1}\Delta_2^{\Delta_2}\Delta_3^{\Delta_3}}\,\GG_{123}\,.
\ee
Absorbing a factor of $\frac{2^{1/2}\Delta_i^{d/4+1/2}}{\pi^{d/4}}$ into $\OO_{\Delta_i}$ so that the two-point function in (\ref{sugra2pt}) is normalized, %then 
(\ref{sugra3pt}) reduces to (\ref{3corr1}) with the coefficient in (\ref{3corr2}).

It is important to note that it is $\Delta^2$ and not the bulk scalar mass-squared, $m^2=\Delta(\Delta-d)$, that appears in the point particle action in (\ref{action}).  If instead we had used $m^2$, then the final result for $C_{123}$ would have come with the extra factor $\left(\frac{(\Delta_1\Delta_2\Delta_3)^2}{\al_1\al_2\al_3\Sigma^{3}}\right)^{d/4}$.

\section{Conserved charges}

In the previous section we have seen that the canonical momenta in (\ref{3pteom}) of the  three geodesics is conserved at the joining point.  The components of the momentum along the transverse directions are constants of the motion, although the component $\Pi_z$ is not.  However, there are other constants of the motion  that end up being conserved by (\ref{3pteom}).

To see how this works, consider the conformal algebra generated by $P_\mu$, $D$, $M_{\mu\nu}$ and $K^\mu$.  Using the convention in \cite{Beisert:2010kp}, their action on a primary operator  $\OO(x)$ is
\be\label{PDKop}
{}[P_\mu,\OO(x)]&=&i\p_\mu \OO(x)\nn\\
{} [D,\OO(x)]&=&i(x^\mu\p_\mu+\Delta)\OO(x)\nn\\
{} [M_{\mu\nu},\OO(x)]&=&i(x_\mu\p_\nu-x_\nu\p_\mu)\OO(x)\nn\\
{} [K^\mu,\OO(x)]&=&i(x^\mu x^\nu\p_\nu+x^\mu\Delta-\half x^\nu x_\nu\p^\mu)\OO(x)\,.
\ee
If we now consider the two-point function $\lb\OO(x)\OO(y)\rb$ with an insertion of  these operators we find
\be\label{consq}
\lb\OO(x)P_\mu\OO(y)\rb&=&2i\,\Delta\frac{x_\mu-y_\mu}{(x-y)^2}\lb\OO(x)\OO(y)\rb\nn\\
\lb\OO(x)D\OO(y)\rb&=&i\,\Delta\frac{x^2-y^2}{(x-y)^2}\lb\OO(x)\OO(y)\rb\nn\\
\lb\OO(x)M_{\mu\nu}\OO(y)\rb&=&i\,\Delta\frac{x_\mu y_\nu-x_\nu y_\mu}{(x-y)^2}\lb\OO(x)\OO(y)\rb\nn\\
\lb\OO(x)K^\mu\OO(y)\rb&=&i\,\Delta\frac{y^\mu x^2-x^\mu y^2}{(x-y)^2}\lb\OO(x)\OO(y)\rb\,.
\ee

We should be able to derive these same coefficients from the path integral.  In $AdS_{d+1}$ the corresponding generators for $SO(2,d)$ are
\be\label{genSO2d}
P_\mu&=&i\,\p_\mu\nn\\
D&=&i\,(x^\nu\p_\nu+z\p_z)\nn\\
M_{\mu\nu}&=&i\,(x_\mu\p_\nu-x_\nu\p_\mu)\nn\\
K^\mu&=&i\,\big(x^\mu (x^\nu\p_\nu+ z\p_z)-\half( x^\nu x_\nu+z^2)\p^\mu\big)\,,
\ee
Generalizing the solution in (\ref{xzsol}) to have arbitrary endpoints
\be
x^\mu(s)&=&R^\mu\tanh(\kappa s)+x_0^\mu\nn\\
z(s)&=&R\sech(s)\,,
\ee
where 
\be\label{Rx0}
R^\mu=\frac{x^\mu-y^\mu}{2}\,,\qquad x_0^\mu=\frac{y^\mu+x^\mu}{2}\,,
\ee
we  replace the  momentum operators $-i\p_\mu$ and $-i\p_z$ with \be\label{momsub}
-i\,\p_\mu&&\!\!\!\!\longrightarrow\ -i\,E^{-1}z^{-2}\dot x_\mu(s)\nn\\
-i\,\p_z&&\!\!\!\!\longrightarrow\ -i\,E^{-1}z^{-2}\dot z(s)\,.
\ee
(The factors of $-i$ on the righthand side are again because $s$ is Euclidean.) We then find 
\be\label{PDKexp}
&&\lb P_\mu\rb=i\Delta \frac{R_\mu}{R^2}\,,\qquad \lb D\rb=i\Delta \frac{x_0^\mu R_\mu}{R^2}\,,\qquad \lb M_{\mu\nu}\rb=-i\Delta \frac{(x_{0\mu}R_\nu-x_{0\nu}R_\mu)}{R^2}\nn\\
&&\qquad\qquad\lb K^\mu\rb=i\Delta\left(\frac{x_0^\mu x_0^\nu R_\nu}{R^2}-\frac{1}{2}\,\frac{R^\mu x_0^2}{R^2}- \frac{1}{2}R^\mu\right),
\ee
which gives
the coefficients in (\ref{consq}) after substituting (\ref{Rx0}).  Note that $\langle K^\mu\rangle=0$ if $x_\mu$ or $y_\mu$ is zero, as expected for a primary operator located at the origin.  

One can also  verify that the classical value for the quadratic Casimir is
\be
-2\lb P_\mu\rb\lb K^\mu\rb+\lb D\rb^2-\frac{1}{2}\lb M_{\mu\nu}\rb\lb M^{\mu\nu}\rb=-\Delta^2\,.
\ee
Substituting the canonical momenta in  (\ref{momsub}) into (\ref{genSO2d}), one finds that the Casimir is
equal to the  momentum squared of the particle in (\ref{momsq}).

If we now turn to three-point functions then we can insert the generators between one operator and the other two.  For example, $D$ can be inserted as
\be
\lb\OO_1(x_1)\OO_2(x_2)D\OO_3(x_3)\rb\,.
\ee
In the path integral this corresponds to taking the conserved value of $D$ on the third trajectory. Using (\ref{Riexp}), (\ref{x-xi}), (\ref{PDKexp}) as well as $x_i=x_{0i}-R_i$, one finds the conserved charge
\be
i\left(\Delta_3-2\,x_{3\mu}\left(\al_1\frac{x_{32}^\mu}{x_{32}^2}+\al_2\frac{x_{31}^\mu}{x_{31}^2}\right)\right)\,.
\ee
It is straightforward to check that this plus the sum of the charges coming from the other two trajectories is zero.  It is also clear that this is consistent with the $D$ commutator in (\ref{PDKop})  when inserted into the correlator for three primary operators.  Analogous expressions can be found for insertions of $K_\mu$.

It will be convenient for later to consider the conserved quantities when the operator positions are shifted by $-x^\mu$, such that the joining point moves to the origin.  In this case, $\langle M_{\mu\nu}\rangle=0$ for all three operators, while the $\langle K_i^\mu\rangle$ are related to $\langle P_i^\mu\rangle$ by
\be\label{KPrel}
\langle K_i^\mu\rangle=-\frac{\al_1\al_2\al_3\Sigma}{F^2}x_{12}^2x_{23}^2x_{31}^2 \langle P_i^\mu\rangle\,,
\ee

\section{Contributions from $S^5$}

In the full string theory with $AdS_5\times S^5$ we also need to include the $S^5$ contribution.  For three-point functions between certain large $J$ states, we can approximate the trajectories of particles on $S^5$ using the peaked harmonics.  We then consider states where the peaks for all three trajectories intersect.  The peaks lie on geodesics, so this would correspond to the three geodesics intersecting at a north and a south pole.

We can parameterize the $S^5$ using the coordinates, $X^I$ where
\be
\sum_{I=1}^6{X^I}^2=1\,.
\ee
For three-point functions, without any loss of generality we can restrict the peaks to lie on the $S^2$ subspace
\be
\sum_{I=1}^3X_I^2=1\,.
\ee
We then choose the three normalized wave-functions
\be
\psi_{J_1}(\vec X)&=&\frac{\sqrt{(J_1+1)(J_1+2)}}{\sqrt{2\pi^3}}(X_1+iX_2)^{J_1}\nn\\
\psi_{J_2}(\vec X)&=&\frac{\sqrt{(J_2+1)(J_2+2)}}{\sqrt{2\pi^3}}(X_1-i(\cos\chi X_2+
\sin\chi X_3))^{J_2}\nn\\
\psi_{J_3}(\vec X)&=&\frac{\sqrt{(J_3+1)(J_3+2)}}{\sqrt{2\pi^3}}(X_1-i(\cos\chi' X_2-
\sin\chi' X_3))^{J_3}\,,
\ee
which are eigenfunctions of the d'Alembertian on $S^5$ with eigenvalues $J_i(J_i+4)$.
If $J_i\gg1$ then the overlap integral
\be
\int d\vec X \delta(1-|X|)\psi_{J_1}(\vec X)\psi_{J_2}(\vec X)\psi_{J_3}(\vec X)
\ee
is dominated near the antipodal intersection  points $X_1=\pm1$.
Hence, we can approximate the overlap as
\be\label{overapp}
\langle \psi_{J_1}\psi_{J_2}\psi_{J_3}\rangle\approx2\int \prod _{I=1}^5dX_I \frac{J_1J_2J_3}{(2\pi^3)^{3/2}}e^{i[(J_1-\cos\chi J_2-\cos\chi' J_3)X_2-(\sin\chi J_2-\sin\chi' J_3)X_2]}e^{-\frac{1}{2}A_{IJ}X_IX_J}\nn\\
\ee
where
\be
A_{22}&=&\sin^2\chi J_2+\sin^2\chi' J_3\qquad A_{33}=J_1+\cos^2\chi J_2+\cos^2\chi' J_3\nn\\ A_{23}&=&A_{32}=\sin\chi\cos\chi J_2-\sin\chi'\cos\chi' J_3\,,
\ee
and
\be
A_{44}=A_{55}=A_{66}=J_1+J_2+J_3\,.
\ee
The overall factor of $2$ in front of the integral in (\ref{overapp}) is because there are two intersection points.
Since the $J_i$ are large, the integrals over $X_2$ and $X_3$ force the angular momentum conservation
\be
J_1-\cos\chi J_2-\cos\chi' J_3=0\,,\qquad \sin\chi J_2-\sin\chi' J_3=0\,,
\ee
which has the solution
\be
\cos\chi=\frac{J_1^2+J_2^2-J_3^2}{2J_1J_2}\qquad \cos\chi'=\frac{J_1^2+J_3^2-J_2^2}{2J_1J_3}\,.
\ee
The Gaussian integrals in (\ref{overapp}) then give
\be
\langle \psi_{J_1}\psi_{J_2}\psi_{J_3}\rangle&\approx&2\frac{(2\pi)^{5/2}J_1J_2J_3}{(2\pi^3)^{3/2}}(\det M)^{-1/2}\nn\\
&=&\frac{1}{2\pi^2}\frac{(J_1J_2J_3)^{3/2}}{(\tilde\al_1\tilde\al_2\tilde\al_3\tilde\Sigma^5)^{1/2}}\,,\ee
where
\be
\tilde\Sigma=\frac{J_1+J_2+J_3}{2},\ \ \tilde\al_1=\frac{J_2+J_3-J_1}{2},\ \ \tilde\al_2=\frac{J_3+J_1-J_2}{2},\ \ \tilde\al_3=\frac{J_1+J_2-J_3}{2}\,.
\ee

In terms of the standard spherical harmonic coefficients, $C^J_{I_1\dots I_J}$ (see \cite{Lee:1998bxa} for definititions),   we have
\be
\langle C^{J_1}C^{J_2}C^{J_3}\rangle&=&(\cos^2\frac{\chi}{2})^{\tilde\al_3}(\cos^2\frac{\chi'}{2})^{\tilde\al_2}(\sin^2\frac{\chi+\chi'}{2})^{\tilde\al_1}\nn\\
&=&\frac{\tilde\al_1^{\tilde\al_1}\tilde\al_2^{\tilde\al_2}\tilde\al_3^{\tilde\al_3}\tilde\Sigma^{\tilde\Sigma}}{J_1^{J_1}J_2^{J_2}J_3^{\tilde J_3}}\,,
\ee
where $\langle C^{J_1}C^{J_2}C^{J_3}\rangle$ is the unique $SO(6)$ invariant made from contracting indices.
Hence, the overlaps can be written as
\be\label{3S5}
\langle \psi_{J_1}\psi_{J_2}\psi_{J_3}\rangle&\approx&\frac{1}{2\pi^2}\frac{(J_1J_2J_3)^{3/2}}{(\tilde\al_1\tilde\al_2\tilde\al_3\tilde\Sigma^5)^{1/2}}\frac{J_1^{J_1}J_2^{J_2}J_3^{\tilde J_3}}{\tilde\al_1^{\tilde\al_1}\tilde\al_2^{\tilde\al_2}\tilde\al_3^{\tilde\al_3}\tilde\Sigma^{\tilde\Sigma}}\langle C^{J_1}C^{J_2}C^{J_3}\rangle\,.
\ee
For chiral primaries where $\Delta_i=J_i$, we can see that except for the prefactors, the contribution of the overlap is the inverse of the contribution from the $AdS$ path integral.

\section{Flat space vertex operators}

The key feature about the intersections of the geodesics in $AdS_5$ and in $S^5$ is that the overlaps occur over a relatively small region of the space-time.  In the $AdS_5$ case, the size of the overlap is more or less the {\it rms} fluctuation of the intersection point $(\vec x,z)$.  For $AdS_5$ this is roughly
\be
\Delta R_{AdS}\sim \left(\frac{\Delta_1\Delta_2\Delta_3}{\al_1\al_2\al_3\Sigma^5}\right)^{1/10}\,.
\ee
For the geodesics on  $S^5$, the interaction region is the size of the 3 wave-function overlap, 
\be
\Delta R_{S}\sim \left(\frac{J_1J_2J_3}{\tilde\al_1\tilde\al_2\tilde\al_3\tilde\Sigma^5}\right)^{1/10}\,.
\ee
Hence, if  $\Delta_i\gg1$ and $J_i\gg1$ such that none of the $\al_i$ and $\tilde\al_i$ are small, then $\Delta R_{AdS}\ll1$ and $\Delta R_{S}\ll1$.  Therefore, over the intersection regions we can ignore the curvatures of $AdS_5$ and $S^5$, as well as the background 5-form flux and approximate the interactions as flat-space interactions.

One important point is that the strings are on-shell.   We saw in section two that at the intersection point there was a conservation of the canonical momentum $\Pi_m$.  If we combine the $AdS_5$ momentum with the momentum on $S^5$, such that
\be
k^M_i=(\vec\Pi_i,\Pi_{zi};\vec J_i)\,,
\ee
then  using the Poincar\'e metric we have
\be\label{k2}
 k_i\cdot k_i=-\Delta_i^2+ J_i^2=-4n\sqrt{\la}\,,
\ee
where $n$ is the string level.  We can then use these $k_i^M$ as the momenta for the string vertex operators (However, we will later find it more convenient to choose the momentum in terms of the charges of the conformal algebra.)  Since the interaction is  centered at a particular point in $AdS_5\times S^5$, we should really be considering wave-packets \cite{Polchinski:1999ry}.  But note that while $\Delta R_{AdS}$ and $\Delta R_S$ are small compared to the radii of curvatures,  they are large compared to the Compton wave-lengths of the string states.  Hence, we can treat these as one would treat  broad wave-packets.  The upshot of this is to replace the usual plane-wave factor of  $(2\pi)^{10}\delta^{10}(k_1+k_2+k_3)$  with overlaps, but to otherwise compute the amplitude as one would for plane-waves, where one uses the central value of the momenta in the vertex operators.

One is now free to choose their favorite prescription for doing string theory three-point functions.  
We will use an NSR prescription and borrow results from \cite{Friedan:1985ge,Koh:1987hm,Kostelecky:1986xg,Polchinski:1998rr}\nocollect{Polchinski:1998rr}, although one could also choose to use the light-cone vertex operators in \cite{Green:1983hw}.  There are seemingly many choices for the vertex operators and we must decide which vertex operator is the correct one to use.    One can decide this by ``flattening" the superconformal algebra.

Let us first show how this works for the conformal algebra.  In $d$ dimensions the conformal algebra is given by 
\begin{eqnarray}\label{confalg}
&&\qquad[D,P_\mu]=iP_\mu\qquad[D,M_{\mu\nu}]=0\qquad[D,K_\mu]=-iK_\mu\nonumber\\
&&[M_{\mu\nu},P_\lambda]=i(\eta_{\mu\la}P_\nu-\eta_{\la\nu}P_{\mu})\qquad
 [M_{\mu\nu},K_\lambda]=i(\eta_{\mu\la}K_\nu-\eta_{\la\nu}K_{\mu})\qquad\nonumber\\
&&\qquad\qquad\qquad[P_\mu,K_\nu]=i(M_{\mu\nu}-\eta_{\mu\nu}D)\,.
\end{eqnarray}
We can put the algebra in the more manifest $SO(2,d)$ form by defining
\be\label{so2,d}
M_{-1\mu}\equiv \sfrac{1}{\sqrt{2}}(\kappa P_\mu-\kappa^{-1}K_\mu)\qquad M_{d\mu}\equiv \sfrac{1}{\sqrt{2}}(\kappa P_\mu+\kappa^{-1}K_\mu)\qquad M_{-1d}\equiv -D\,,
\ee
where $\kappa$ is arbitrary.  In this convention the casimir becomes $-\half M_{rs}M^{rs}=-\Delta^2$, $r,s=-1\dots d$.
If we now define
\be\label{P_m}
\hat P_m\equiv\eps M_{-1m}\qquad m=0\dots d\,,
\ee
and take the limit $ \eps\to0$, then (\ref{confalg}) reduces to the Poincare algebra in $d+1$ dimensions.   If we set $d=4$ and include the $SO(6)$ $R$-symmetry generators $R_{IJ}$, then we can combine the two Casimirs into
\be
-\half M_{rs}M^{rs}+\half R_{IJ}R^{IJ}=-\Delta^2+J^2\,.
\ee

By doing an overall shift of  the operator positions such that all $\langle M_{\mu\nu}\rangle=0$,
then $\langle K^\mu_i\rangle$ is related to $\langle P^\mu_i\rangle$ by (\ref{KPrel}).  If one then chooses $\kappa$ to be,
\be
\kappa =\frac{\sqrt{\al_1\al_2\al_3\Sigma}}{F}|x_{12}||x_{23}||x_{31}|\,,
\ee
then the only nonzero components in (\ref{so2,d}) are $\langle M_{-1m}\rangle$  for all three operators.  Assuming $\Delta^2\gg1$, we can  then identify the full 10-dimensional flat-space momentum as
\be\label{10dmom}
k^M=(M_{-1m};R_{J6})\,.
\ee
 This satisfies the on-shell condition in (\ref{k2}).

Now consider the $\NN=4$ superconformal algebra in 4 dimensions.  In addition to (\ref{confalg}), the superconformal algebra includes the anticommutators
\begin{eqnarray}\label{QQ}
\{Q_{\al\,a},\tQ^b_{\dda}\}&=&\s^\mu_{\al\dda}{\de_{a}}^bP_\mu\,,\qquad\{S^a_{\al},\tS_{\dda\,b}\}=\s^\mu_{\al\dda}{\de^a}_bK_\mu\qquad{ a,b=1\dots 4,\ \ \al,\dda=1,2}\nn\\
\{Q_{\al\,a},S^b_{\beta}\}&=&-\sfrac{i}{2}\,\varepsilon_{\al\beta}{{\gamma^{IJ}}_a}^ bR_{IJ}+\sfrac12\,\s^{\mu\nu}_{\al\beta}{\de_a}^bM_{\mu\nu}-\sfrac{1}{2}\,\varepsilon_{\al\beta}{\de_a}^b\,D\nonumber\\
\{ \tQ^a_{\dda},\tS_{\ddb\,b}\}&=&+\sfrac{i}{2}\,\varepsilon_{\dda\ddb}{\gamma^{IJ\,a}}_{b}\,R_{IJ}+\sfrac{1}{2}\,\s^{\mu\nu}_{\dda\ddb}{\de^a}_bM_{\mu\nu}-\sfrac{1}{2}\,\varepsilon_{\dda\ddb}{\de^a}_bD\nonumber\\
\{Q_{\al\,a},\tS_{\ddb\,b}\}&=&\{\tQ^a_{\dda},S^b_{\beta}\}=\{Q_{\al\,a},Q_{\al\,b}\}=
\{\tQ^a_{\dda},\tQ^b_{\dda}\}=\{S^a_{\al},S^b_{\al}\}=
\{\tS_{\dda\,a},\tS_{\dda\,b}\}=0\,.\nn
\end{eqnarray}
We can put the algebra into a more symmetric form 
by defining
\be
Q^1_{\dot a a}&\equiv &(\kappa^{1/2}Q_{\al a},\kappa^{-1/2}\tilde S_{\dot\al a})\nn\\
Q^{2,\dot a a}&\equiv &(\kappa^{-1/2}\eps^{\al\beta}S^a_{\beta},\kappa^{1/2}\eps^{\dot\al\dot\beta}\tilde Q^a_{\dot\beta})
\ee
where the lowered $\dot a$ is an $SO(2,4)$ spinor index and the raised $\dot a$ is an index for the other spinor representation.  The anti-commutators now read
\be
{}&&\{Q^1_{\dot a a},Q^{2,\dot bb}\}=\sfrac{1}{2}\,{\delta_a}^bM_{mn}{{\gamma^{mn}}_{\dot a}}^{\dot b}-\sfrac{i}{2}\,{\delta_{\dot a}}^{\dot b}R_{IJ}{{\gamma^{IJ}}_a}^ b\nn\\
&& \{Q^1_{\dot a a},Q^1_{\dot b b}\}=\{Q^{2,\dot aa},Q^{2,\dot bb}\}=0\,.
\ee
We then define  two  sets of  supercharges
\be\label{QLQRdef}
 Q^{L,R}_A=Q^1_{\dot a a}\pm\gamma^{-1}_{\dot b \dot a}\gamma^{\ 6}_{ba}Q^{2,\dot bb}\,,
\ee
and their rescaled versions, $ \hat Q^{L,R}_A\equiv \sqrt{\eps}\, Q^{L,R}_A$,
where we have chosen particular directions in the 6-dimensional embedding spaces of  $AdS_5$ and $S^5$.  The spinor index $A$ is the  tensor product of the two lowered 6-dimensional spinor indices.  As in (\ref{P_m}) we also choose the five additional rescaled momentum components
\be
 \qquad \hat P_{J+4}=-i\,\eps R_{J6}\,.
\ee
Again taking the limit $\eps\to0$, the anticommutators in (\ref{QQ}) become
\be
\{\hat Q^{L,R}_A,\hat Q^{L,R}_B\}=-2(\mbox{P}_+\Gamma^M\CC)_{AB}\hat P_M\,,\qquad\{\hat Q^{L}_A,\hat Q^{R}_B\}=0
\ee
where  $\Gamma^M$ are  ten-dimensional Dirac matrices, $\CC$ is the charge conjugation matrix and $\mbox{P}_+$ is the positive chirality projector.
These anticommutators are the fermionic part of the ten-dimensional  type IIB super-Poincar\'e algebra.  Using the identification of the flat-space momentum in (\ref{10dmom}),  we see that (\ref{QLQRdef}) are the corresponding  flat-space supergenerators.

We can now use the supergenerators to determine the vertex operators.  For a primary operator $\OO(x)$ we have that  $S^a_\al\OO(0)=\tS_{\dda\,a}\OO(0)=0$ and so $K^\mu\OO(0)=0$.  In the three-point function, if we place one of the operators at  $x^\mu=0$ and move the other two operators far away from the origin, then for any finite choice of $\kappa$ we have $M_{-1\mu}=M_{4\mu}=0$, $M_{-1 4}=-i\Delta$.  The   space-time directions are then transverse to the particle's momentum in $AdS_5$.  Inspecting (\ref{QLQRdef}), we see that in the flat-space limit the annihilation of the primary operator by the superconformal charges corresponds  to 
\be\label{twistcond}
Q^L_{\al a}=Q^R_{\al a}\,,\qquad Q^L_{\dot\al a}=-Q^R_{\dot\al a}
\ee
 for these operators, where we have used the explicit four-dimensional space-time spinor indices.   In the more general case where the joining point is shifted to $x^\mu=0$, the $M_{-1\mu}$ are no longer   zero in general.  However, the primary operator condition is still defined by the directions transverse to the particle momentum in $AdS_5$.  In other words, we still use (\ref{twistcond}), but now the four-dimensional spinor indices correspond to the directions transverse to the particle momentum in $AdS_5$ and not necessarily the space-time directions.
 
Therefore, we choose our vertex operators $V(k)$ such that
\be\label{twistcond2}
Q^L_{\al a}V(k)=Q^R_{\al a}V(k)\,,\qquad Q^L_{\dot\al a}V(k)=-Q^R_{\dot\al a}V(k)
\ee
where the spinor refer to the directions transverse to $k$.
 It is  clear that this can lead to mixing between NS-NS and R-R modes, since schematically we have that
\be\label{QLQRNSR}&&Q_L(|NS\rangle\otimes|NS\rangle+|R\rangle\otimes|R\rangle)=|R\rangle\otimes|NS\rangle+|NS\rangle\otimes R\rangle\nn\\
&&Q_R(|NS\rangle\otimes|NS\rangle+|R\rangle\otimes|R\rangle)=|NS\rangle\otimes|R\rangle+|R\rangle\otimes NS\rangle
\ee
and so  requires a mixture of both sets of fields in order that the right-hand sides in (\ref{QLQRNSR}) are equal, or  equal up to a sign.   

Note that the primary operator condition in (\ref{twistcond2})  only applies to the leading order approximation.  At the next order in $1/\sqrt{\la}$ there is a deformation of the super Poincar\'e algebra, which  results in  a mixing of vertex operators satisfying (\ref{twistcond2}) with other vertex operators.

To solve (\ref{twistcond2}) we find it convenient to consider the twisted vertex operators,
\be
V_T(k)=e^{\pi i(M^R_{0'1'}+M^R_{2'3'})}V(k)\,,
\ee
where the $\mu'$ components are transverse to $k^M$ and $M^R_{\mu'\nu'}$ is a generator of rotations for the right-movers.  Hence, the twisted vertex operators satisfy
\be
Q^L_{A}V_T(k)=Q^R_{A}V_T(k)
\ee
for all components.

\subsection{Level $0$ }
Before tackling the more difficult case of  vertex operators at level 1, let us consider those at level 0.  In this case, the corresponding operators are chiral primaries.
The momentum of one of these states satisfies $k^2=0$, corresponding to $\Delta^2= J^2$. In the NS-NS sector a generic vertex operator in the $(-1,-1)$ picture is given by \cite{Friedan:1985ge}
\be
V^{-1,-1}(z,\bar z)=g_s\ve_{MN}\psi^M\tilde\psi^Ne^{-\phi-\tilde\phi}e^{ik\cdot X}\qquad
k^M\ve_{MN}=k^N\ve_{MN}=0\,,
\ee
where $\phi$ and $\tilde\phi$ are the left and right  superconformal ghost fields.
In the AdS/CFT dictionary we have that $g_s=1/N$ and $\al'=\sqrt{\la}$, but we will leave these in their familiar string-theory form. 
To compute certain three-point functions we will also need the vertex in the $(0,0)$ picture, \be
V^{0,0}(z,\bar z)=-g_s\ve_{MN}\sfrac{2}{\al'}\big(i\p X^M+\sfrac{\al'}{2}k\cdot\psi\psi^M)\big(i\bar\p X^M+\sfrac{\al'}{2}k\cdot\tilde\psi\tilde\psi^M)e^{ik\cdot X}\,.
\ee
 In the R-R sector  a generic vertex operator is given by  \cite{Friedan:1985ge}
\be
V^{-1/2,-1/2}(z,\bar z)=g_s\left(\sfrac{\al'}{2}\right)^{1/2}t^{AB}\tilde\Theta_A\Theta_B\,e^{-\half\phi-\half\tilde\phi}e^{ik\cdot X}\,,\quad{t\,\slash\!\!\! k=0}\,,
\ee
where $\Theta_A$ and $\tilde\Theta_B$ are the 16 component left and right twist fields.

The supercharges $Q^L_A$ and $Q^R_A$ are made from the zero-momentum left and right vertex for a Ramond field  \cite{Friedan:1985ge}, where for  the $+1/2$ picture of $Q^L$ we have
 \be
 Q_{A}^{L(+1/2)}=\left(\sfrac{2}{\al'}\right)^{3/4}\oint \frac{dz}{2\pi i}\,i\p X_M\Gamma_{AB}^M\Theta^B\e^{+\half\phi}\,,
 \ee
 while for the  $-1/2$ picture of $Q_R$ we have
\be
Q_{A}^{R(-1/2)}=\left(\sfrac{2}{\al'}\right)^{1/4}\oint \frac{d\bar z}{2\pi i}\tilde\Theta_Ae^{-\half\tilde\phi}\,.
\ee
Acting with the left supercharge on the NS-NS vertex we find
\be\label{QLNSNS}
Q_{A}^{L(+1/2)}V^{-1,-1}(z,\bar z)=\sfrac{1}{\sqrt{2}}g_s\ve_{MN}\left(\sfrac{\al'}{2}\right)^{1/4}{(\slash\!\!\! k\Gamma^M)_A}^B\theta_B\tilde\psi^Ne^{-\half\phi-\tilde\phi}e^{ikX}
\ee
while acting with the right supercharge on the R-R vertex we get
\be\label{QRRR}
Q_{A}^{R(-1/2)}V^{-1/2,-1/2}(z,\bar z)=\sfrac{1}{\sqrt{2}}g_st^{CB}\left(\sfrac{\al'}{2}\right)^{1/4}{(\Gamma_N\CC)_{AC}}\Theta_B\tilde\psi^Ne^{-\half\phi-\tilde\phi}e^{ikX}\,.
\ee
To find the twisted vertex, we set (\ref{QLNSNS}) equal to (\ref{QRRR}), resulting in the relation
\be\label{n=0rel}
\ve^{MN}_T{(\slash\!\!\! k\Gamma_M)_A}^B=t^{CB}_T{(\Gamma^N\CC)_{AC}}\,.
 \ee
This has one scalar  solution, up to an overall normalization,
\be\label{soln=0}
\ve^{MN}_T=\eta^{MN}-\frac{k^M\bar k^N+k^N\bar k^M}{k\cdot\bar k}\,,\qquad\qquad t_T^{AB}=(\CC\slash\!\!\! k)^{AB}\,,
\ee
where $\bar k^N$ is an arbitrary light-like vector satisfying $k\cdot\bar k\ne 0$.  Actually, (\ref{soln=0}) does not exactly satisfy (\ref{n=0rel}).  But we observe that the righthand side of (\ref{n=0rel}) with the expression for $t^{AB}$ in (\ref{soln=0}) satisfies $t^{CB}{(\Gamma_N\CC)_{AC}}k^N=0$.  Hence, we can replace $\Gamma_N$ here with $\Gamma_N-({\slash\!\!\! k \bar k_N+\slash\!\!\! \bar k k_N})/({k\cdot\bar k})$, which modifies the vertex in (\ref{QRRR}) by a spurious term.  After this replacement, (\ref{soln=0}) satisfies (\ref{n=0rel}).

The twisted vertex operator is a linear combination of a dilaton and axion vertex operator.  If we now untwist the operator,  the relation of the untwisted polarization to the twisted for the NS-NS vertex is
\be
\ve^{MN}&=&(-1)^{\s(N)}\ve^{MN}_T\,
\ee
where
\be\label{sdef}
\s(N)&=&1\qquad N=0\dots 4\nn\\
\s(N)&=&0\qquad N=5\dots 9\,.
\ee
This corresponds to a graviton state in the full 10-dimensional space-time.  In the R-R sector, $t_T^{AB}$ is replaced with 
\be
t^{AB}=(\CC\, (i\,\Gamma^{0'}\Gamma^{1'}\Gamma^{2'}\Gamma^{3'})\slk)^{AB}\,,
\ee
 Hence, this state corresponds to a component of the self-dual tensor.
Thus, the vertex operator is consistent with the supergravity computation of  \cite{Lee:1998bxa}, where it was shown that the primary state is dual to a linear combination of the graviton and the 4-form tensor field.

\subsection{Level 1}

At level 1, where $k^2=-4/\al'$,  there are two types of NS-NS vertices.  In the $(-1,-1)$ picture these have the form \cite{Koh:1987hm}
\be
V_1^{(-1,-1)}(z,\bar z)&=&g_s\left(\sfrac{2}{\al'}\right)\eps_{MN;\tM\tN}\,\psi^M(z)\p X^N \tilde\psi^{\tM}(\bar z)\bar\p X^{\tN} \,e^{ik\cdot X}e^{-\phi-\tilde\phi}\,,\nn\\
V_2^{(-1,-1)}(z,\bar z)&=&g_s\,\al_{MNL;\tM\tN\tL}\,\psi^M(z) \psi^N(z)\psi^L(z)\,\psi^\tM(\bar z) \psi^\tN(\bar z)\psi^\tL(\bar z)\, e^{ik\cdot X}e^{-\phi-\tilde\phi}\,.\nn\\
\ee
The first two and last two indices of $\eps_{MN;\tM\tN}$ are symmetric and traceless, while the first three and last three indices of $\al_{MNL;\tM\tN\tL}$ are antisymmetric.  The contraction of any index in $\eps_{MN;\tM\tN}$ or $\al_{MNL;\tM\tN\tL}$ with $k^M$ is zero for physical states.  There are $44\times 44$ independent polarizations of the first type and $84\times 84$ of the second type.
However, only one of each type is a Lorentz scalar.   

 For the first case the (unnormalized) polarization for the twisted vertex is
\be
\eps_T^{MN;\tM\tN}=
(\sfrac12(\hat\eta^{M\tM}\hat\eta^{N\tN}+\hat\eta^{M\tN}\hat\eta^{N\tM})-\sfrac{1}{9}\,\hat\eta^{MN}\hat\eta^{\tM\tN})\,,
\ee
where $\hat\eta^{MN}\equiv\eta^{MN}-\frac{k^M k^N}{k^2}$.  For the second we have
\be
\al_T^{MNL;\tM\tN\tL}=
\sfrac{1}{(3!)^2}(\hat\eta^{M\tM}\hat\eta^{N\tN}\hat\eta^{L\tL}-\mbox{\ perms})\,.
\ee
 The vertex operators in the $(0,0)$ picture are given by
\be
V_{1T}^{(0,0)}&=&-g_s\left(\sfrac{2}{\al'}\right)^2\eps_{T\,MN;\tM\tN}\big(\p X^M(i\,\p X^N+\sfrac{\al'}{2}k\!\cdot\!\psi \,\psi^N)-i\,\p\psi^M\psi^N\big)\nn\\
&&\qquad\qquad\qquad\qquad\times\big(\bar\p X^\tM(i\,\bar\p X^\tN+\sfrac{\al'}{2}\,k\!\cdot\!\tilde\psi\, \tilde\psi^\tN)-i\,\bar\p\tilde\psi^\mu\tilde\psi^\nu\big)e^{ik\cdot X}\,,\nn\\
V_{2T}^{(0,0)}(z,\bar z)&=&-\,g_s\left(\sfrac{2}{\al'}\right)\,\al_{T\,MNL;\tM\tN\tL}\, (3\,i\,\p X^M+\sfrac{\al'}{2}k\!\cdot\!\psi \,\psi^M)\psi^N\psi^L\,\nn\\
 &&\qquad\qquad\qquad\qquad\times(3\,i\,\bar\p X^\tM+\sfrac{\al'}{2}\,k\!\cdot\!\tilde\psi\, \tilde\psi^\tM)\psi^\tN\psi^\tL\, e^{ik\cdot X}\,.\nn\\
\ee

The left-moving part of the level one R-R vertex  in the $(-1/2,-1/2)$ picture has the form  \cite{Koh:1987hm} 
\be
\big(v_{MA}\,i\,\p X^M\Theta^A+\rho_{M A}\psi^M(\slash\!\!\!\!\psi\Theta)^{ A}\big)e^{-\phi/2}\,,
\ee
where BRS invariance imposes the conditions
\be
\slash\!\!\!v&=&\left(\sfrac{\al'}{2}\right)(v\cdot k)\slash\!\!\!k\nn\\
\rho^M&=&-\sfrac18\left(\sfrac{\al'}{2}\right)v^M\slash\!\!\!k+\sfrac{1}{36}\left(\sfrac{\al'}{2}\right)(v\cdot k)\Gamma^M\,.
\ee
This still leaves 16 spurious modes, but they can be removed with the further condition
\be
\slash\!\!\! v=0\,,
\ee
which reduces the BRS conditions to 
\be
v\cdot k=0\,,\qquad \rho^M=-\sfrac18v^M\slash\!\!\!k\,.
\ee
Tensoring this with the corresponding right-moving part we can construct one scalar, whose unnormalized vertex is given by
\be
&&V_{3\,T}^{(-1/2,-1/2)}(z,\bar z)=g_s\left(\sfrac{2}{\al'}\right)^{1/2}\nn\\
&&\qquad 
\times\big(i\,\bar\p X^M\tilde\Theta\!-\!\sfrac18\left(\sfrac{\al'}{2}\right)\tilde\psi^M(\slash\!\!\!k\,\slash\!\!\!\!\tilde\psi\tilde\Theta)\big)^At_{TMA;NB}\big(i\,\p X^N\Theta\!-\!\sfrac18\left(\sfrac{\al'}{2}\right)\psi^N(\slash\!\!\!k\,\slash\!\!\!\!\psi\Theta)\big)^B\nn\\
&&\qquad\qquad\qquad\qquad\qquad\qquad\times\, e^{ikX}e^{-\phi/2-\tilde\phi/2}\,,
\ee
where
\be
t_{TMA;NB}=\left(\CC\,\slk(\hat\eta_{MN}-\sfrac19\hat\Gamma_M\hat\Gamma_N)\right)_{AB}
\ee
and
 $\hat\Gamma^M=\Gamma^M-{\slk k^M}/{k^2}$.

Acting with $Q_{A}^{L(+1/2)}$ on $V_{1T}^{-1,-1}$ and $V_{2T}^{-1,-1}$, we end up with the vertex operators in the R-NS sector   \cite{Koh:1987hm},
\be\label{L1QL}
Q_{A}^{L(+1/2)}V_{1T}^{-1,-1}&=&-i\,g_s\sfrac{1}{\sqrt{2}}\left(\sfrac{2}{\al'}\right)^{3/4}\eps_{T\,MN;\tM\tN}(\slk\hat\Gamma^M)_{AB}\,(i\,\p X^N\Theta^B\!-\!\sfrac18\left(\sfrac{\al'}{2}\right)\psi^N(\slash\!\!\!k\,\slash\!\!\!\!\psi\Theta)^B) \nn\\
&&\qquad\qquad\qquad\times\,\tilde\psi^{\tM}\bar\p X^{\tN}e^{-\phi/2-\tilde\phi} \,e^{ik\cdot X}\nn\\
&=&-i\,g_s\sfrac{1}{\sqrt{2}}
\left(\sfrac{2}{\al'}\right)^{3/4}(\slk\hat\Gamma^M)_{AB}\,(i\,\p X^N\Theta^B\!-\!\sfrac18\left(\sfrac{\al'}{2}\right)\psi^N(\slash\!\!\!k\,\slash\!\!\!\!\psi\Theta)^B) \nn\\
&&\qquad\qquad\qquad\times\,\left(\sfrac12\tilde\psi_{\{M,}\bar\p X_{N\}}-\sfrac19\hat\eta_{MN}\tilde\psi\cdot\bar\p X\right)e^{-\phi/2-\tilde\phi} \,e^{ik\cdot X}\nn\\
Q_{A}^{L(+1/2)}V_{2\,T}^{-1,-1}&=&\,g_s\sfrac{1}{2\sqrt{2}}\left(\sfrac{2}{\al'}\right)^{3/4}\al_{MNL;\tM\tN\tL}\left(\hat\Gamma_P\hat\Gamma^{MNL}+\sfrac13\hat\Gamma^{MNL}\hat\Gamma_P\right)_{AB}\, \nn\\
&&\qquad\times(i\,\p X^P\Theta^B\!-\!\sfrac18\left(\sfrac{\al'}{2}\right)\psi^P(\slash\!\!\!k\,\slash\!\!\!\!\psi\Theta)^B)\,\tilde\psi^{\tM}\tilde\psi^{\tN}\tilde\psi^{\tL}e^{-\phi/2-\tilde\phi} \,e^{ik\cdot X}\nn\\
&=&\,g_s\sfrac{1}{2\sqrt{2}}
\left(\sfrac{2}{\al'}\right)^{3/4}\left(\hat\Gamma_{\tN\tL}\left(\hat\eta_{P\tM}-\sfrac19\hat\Gamma_\tM\hat\Gamma_P\right)\right)_{AB}\, \nn\\
&&\qquad\times(i\,\p X^P\Theta^B\!-\!\sfrac18\left(\sfrac{\al'}{2}\right)\psi^P(\slash\!\!\!k\,\slash\!\!\!\!\psi\Theta)^B)\,\sfrac{1}{3!}\tilde\psi^{[\tM}\tilde\psi^{\tN}\tilde\psi^{\tL]}e^{-\phi/2-\tilde\phi} \,e^{ik\cdot X}\,,\nn\\
\ee
up to spurious terms.  Likewise, $Q_{A}^{R(-1/2)}$ acting on $V_{3\,T}^{-1/2,-1/2}$ gives the R-NS vertex
\be\label{L1QR}
&&Q_{A}^{R(-1/2)}V_{3\,T}^{-1/2,-1/2}=-g_s\sfrac{1}{\sqrt{2}}\left(\sfrac{2}{\al'}\right)^{3/4}
\left(i\,\p X^N\Theta^B\!-\!\sfrac18\left(\sfrac{\al'}{2}\right)\psi^N(\slash\!\!\!k\,\slash\!\!\!\!\psi\Theta)^B\right)\nn\\
&&\qquad \ \times\left(\left(\sfrac{i}{2}\,\bar\p X^{\{M,}\tilde\psi^{\tN\}}\hat\Gamma_\tN-\sfrac12\left(\sfrac{\al'}{2}\right)\sfrac{1}{3!}\tilde\psi^{[M}\tilde\psi^\tN\tilde\psi^{\tL]}\hat\Gamma_{\tN\tL}\slk\right)\slk\left(\hat\eta_{MN}-\sfrac19\hat\Gamma_M\hat\Gamma_N\right)\right)_{AB}\nn\\
&&\qquad\qquad\qquad\qquad\qquad\qquad\qquad\times\, e^{ikX}e^{-\phi/2-\tilde\phi}\,,
\ee
again up to spurious terms.  Comparing the expressions in (\ref{L1QL}) and (\ref{L1QR}) and using that $-k^2=4/\al'$, we find that the  linear combination
\be\label{twvert}
V_T=\kappa\left(
V_{1T}+
V_{2\,T}- 
V_{3\,T}\right)
\ee
satisfies the $Q^L=Q^R$ condition. The normalization constant $\kappa$ is given by
\be
\kappa=\left(44+ 84+\sfrac{\al'}{2}(-k^2)\cdot64\right)^{-1/2}=\sfrac{1}{16}\,.
\ee

Untwisting the vertex operators we find for the NS-NS polarizations
\be
\ve_{MN;\tM\tN}&=&(-1)^{\s(\tM)+\s(\tN)}\ve_T^{MN;\tM\tN}\nn\\
\al^{MNL;\tM\tN\tL}&=&(-1)^{\s(\tM)+\s(\tN)+\s(\tL)}\al_T^{MNL;\tM\tN\tL}\,,
\ee
where the $\s(\tM)$ are defined in (\ref{sdef}).
For the R-R polarization we have
\be\label{RRpol}
t_{MA;NB}=(-1)^{\s(M)}\left(\CC(i\,\Gamma^{0'}\Gamma^{1'}\Gamma^{2'}\Gamma^{3'})\,\slk\,(\hat\eta_{MN}-\sfrac19\hat\Gamma_M\hat\Gamma_N)\right)_{AB}
\ee

\section{Comments on three-point string vertex amplitudes}

We now have all of the ingredients to compute the coupling $\GG_{123}$ that appears in  (\ref{3corr2}) for three level-one primary operators.  Including the $S^5$ wave-function overlaps, we have
\be\label{coupling}
\GG_{123}= \frac{8\pi}{g_s^2\al'}\,\langle V(k_1)V(k_2)V(k_3)\rangle\ \langle \psi_{J_1}\psi_{J_2}\psi_{J_3}\rangle
\ee
where the $V(k_i)$ are the untwisted versions of (\ref{twvert}).  The factor ${8\pi}/{g_s^2\al'}$ is the standard normalization that is fixed by the massless couplings \cite{Polchinski:1998rr}.  We are especially interested in the case where $ J_i\ll \Delta_i$, which is a situation close to having three Konishi operators.  Even in this case, it is still possible to choose the charges such that $1\ll J_i\ll \Delta_i $, so that the flat-space limit is still valid.  

The combinatorics for three level-one vertex operators in type IIB string theory are rather nasty and  the result of the computation will be deferred to a later publication \cite{To_appear}.  However,  even without doing a direct calculation there is much that we can infer about   $\GG_{123}$, and hence the  constant $C_{123}$ in  (\ref{3corr2}).    For one thing,  
we can  see that in the limit $J_i\ll\Delta_i$, the $R$-charge has very little influence on the vertex operators and hence the string three-point functions.  In fact, in this limit we can effectively replace the R-R polarization in  (\ref{RRpol}) with
\be
t_{MA;NB}\approx(-1)^{\s(M)}\Delta\left(\CC\,\Gamma^{0}\Gamma^{1}\Gamma^{2}\Gamma^{3}\Gamma^4\,(\hat\eta_{MN}-\sfrac19\hat\Gamma_M\hat\Gamma_N)\right)_{AB}\,.
\ee
Hence, the $R$-charges only appear in the wave-function overlaps.  It is therefore tempting for level one and higher states to simply set $J_i=0$ and assume constant wave-functions on $S_5$.  The overlap is then $ \langle \psi_{J_1}\psi_{J_2}\psi_{J_3}\rangle=\pi^{-3/2}$ and  (\ref{coupling}) can then be used to find  the coupling for three Konishi operators.

One can also see that if  $J_i\ll \Delta_i$, then
\be
\langle V(k_1)V(k_2)V(k_3)\rangle\sim g_s^3\,.
\ee
  Hence, the result of the string amplitude is of order $\sqrt{\la}/N$.  When $J_i\ll\Delta_i$, the prefactors  in (\ref{3corr2}) win out over the $S^5$ overlaps in  ({\ref{coupling}), leading to an exponential suppression in $\Delta_i$.  One  then finds
\be\label{Capprox}
C_{123}\sim \frac{1}{N}\exp\left(-3(\la)^{1/4}\log(4/3)\right)\,,
\ee
giving an exponential suppression of $C_{123}$ at large coupling.

\section{Discussion}

In this paper we have mapped out a strategy for computing the three-point functions for short operators that are not chiral primaries.  The key idea is to realize that  one can use flat-space vertex operators in computing the couplings.  In principle, one can also compute $\al'$ corrections by including the curvature terms perturbatively.

In the limit where $\Delta_i\gg J_i$, the only dependence on the $R$-charges is in the wave-function overlaps.  Assuming that one can continue to use the flat-space limit, then one can find the three-point function for three Konishi operators at strong coupling.  Our hope is that eventually it will be possible to compute the coupling using the underlying integrability of $\NN=$ SYM, as it is now possible to compute the dimension of the Konishi operator at any coupling \cite{Gromov:2009zb,Frolov:2010wt} using the ideas in \cite{Gromov:2009tv,Arutyunov:2009zu,Bombardelli:2009ns,Gromov:2009bc,Arutyunov:2009ur}.   These numerical computations were shown to approach the string prediction of \cite{Gubser:1998bc}, as well as the leading $\al'$ correction in \cite{Roiban:2009aa,Gromov:2011de,Roiban:2011fe}.  The recent promising developments using integrability for three-point functions \cite{Escobedo:2010xs,Escobedo:2011xw,Gromov:2011jh,Foda:2011rr,Gromov:2012uv,Foda:2012wf,Bissi:2011ha,Georgiou:2012zj,Gromov:2012vu,Kostov:2012jr,Serban:2012dr,Kostov:2012yq}\nocollect{Bissi:2011ha}\ bode well for further progress in this area.

Besides three short operators at level one, it would also be interesting to investigate the case where two operators are chiral primaries and one is dual to a massive short string state.  It would also be interesting to consider    three-point functions for two heavy semiclassical states and one short state, perhaps using the ideas  in  \cite{Zarembo:2010rr,Costa:2010rz}.

%%%%%%%%%%%%%%%%%%%%%%%%%%%%%%%%
\subsection*{Acknowledgments}
I thank P. Di Vecchia, T. Klose, L. Rastelli, R. Russo and A. Tseytlin for discussions.  This  research is supported in part by
Vetenskapsr\aa det under grant \#2009-4092.  I thank the
CTP at MIT  and Nordita for kind
hospitality  during the course of this work. 
%%%%%%%%%%%%%%%%%%%%%%%%%%%%%%%%%%

\footnotesize
\bibliographystyle{JHEP}
\bibliography{references}

\end{document}